\documentclass[aps,prl,twocolumn,citeautoscript]{revtex4}
\setcitestyle{super}
\usepackage{amsmath}
\usepackage{amsfonts}
\usepackage{amssymb}
\usepackage{graphicx}
\usepackage{float} 
\usepackage[caption=false]{subfig}
\usepackage[usenames,dvipsnames,svgnames,table]{xcolor} 
\usepackage[normalem]{ulem} 
\usepackage[makeroom]{cancel} 
\usepackage[hidelinks]{hyperref} 

\newcommand{\Tstart}{0.07 U_0}
\newcommand{\Tstop}{0.01 U_0}
\newcommand{\dT}{-0.0002 U_0}
\newcommand{\Tc}{0.056 U_0}
\newcommand{\dopingGlobulePercent}{1.0} 
\newcommand{\dopingNetPercent}{1.9} 

\begin{document}

\title{Modeling of networks and globules of charged domain walls observed in pump and pulse induced states.}

\author{Petr Karpov}
\email{karpov.petr@gmail.com}
\affiliation{National University of Science and Technology ``MISiS'', Moscow, Russia}
\author{Serguei Brazovskii}
\affiliation{National University of Science and Technology ``MISiS'', Moscow, Russia}
\affiliation{CNRS UMR 8626 LPTMS, University of Paris-Sud, University of Paris-Saclay, Orsay, France}
\affiliation{Jozef Stefan Institute, Jamova 39, SI-1000 Ljubljana, Slovenia}

\date{September 18, 2017}

\begin{abstract}
Experiments on optical and STM injection of carriers in layered $\mathrm{MX_2}$ materials revealed the formation of nanoscale patterns with networks and globules of domain walls. This is thought to be responsible for the metallization transition of the Mott insulator and for stabilization of a ``hidden'' state.
In response, here we present studies of the classical charged lattice gas model emulating the superlattice of polarons ubiquitous to the material of choice $1T-\mathrm{TaS_2}$.
The injection pulse was simulated by introducing a small random concentration of voids which subsequent evolution was followed by means of Monte Carlo cooling. Below the detected phase transition, the voids gradually coalesce into domain walls forming locally connected globules and then the global network leading to a mosaic fragmentation into domains with different degenerate ground states. The obtained patterns closely resemble the experimental STM visualizations.
The surprising aggregation of charged voids is understood by fractionalization of their charges across the walls' lines.
\end{abstract}

\maketitle

\section{Introduction}

Major anticipations for the post-silicon electronics are related to materials which demonstrate a layered structure with a possibility for exfoliation down to a few and even a single atomic layer, akin to the graphene. The latest attention was paid to oxides and particularly di-halcogenides of transition metals $MX_2$ with $M$ = Nb, Ta, Ti and $X$ = S, Se, see e.g. \cite{Manzeli:2017,Rossnagel:2011} for reviews.

These materials show a very rich phase diagram spanning from unconventional insulators of the so called Peierls and Mott types to the superconductivity \cite{Sipos:2008}. The transformations among these phases involve formation of superstructures like several types of so called charge density waves (CDW) and/or of hierarchical polaronic crystals.
Recent studies of these materials fruitfully overlapped with another new wave in solid state physics. This is the science of controlled transformations of electronic states or even of whole phases by impacts of strong electric fields and/or the fast optical pumping. A super goal is to attend ``hidden'' states which are inaccessible and even unknown under equilibrium conditions. In relation to this article subjects, the success came recently from observations of ultrafast (at the scale of picoseconds) switching by means of optical \cite{Stojchevska:2014,Gerasimenko:2017} and voltage
\cite{Vaskivskyi:2016,Yoshida:2016} pulses, as well by local manipulations \cite{Ma:2016,Cho:2016}. The registered ultrafast switching is already discussed as a way for new types of RAM design, see \cite{Svetin:2016} and rfs. therein.

Most challenging and inspiring observations have been done in studies by the scanning tunneling microscopy (STM) and spectrometry (STS) \cite{Ma:2016,Cho:2016,Cho:2017,Gerasimenko:2017}. They have shown that the switching from an insulating to a conducting state proceeds via creation of local globules or extended networks of domain walls enforcing fragmentation of the insulating electronic crystal into a conducting mosaics of domains with different multiply degenerate ground states.

Most important observations have been done upon very popular nowadays layered material $1T-\mathrm{TaS_2}$ which is a still enigmatic ``polaronic Wigner-crystalline Mott insulator''. The rich phase diagram of $1T-\mathrm{TaS_2}$ includes such states as incommensurate, nearly commensurate, and commensurate CDWs which unusually support also the Mott insulator state for a subset of electrons. Recently, new long-lived metastable phases have been discovered: a ``hidden'' state created by laser \cite{Stojchevska:2014, Gerasimenko:2017} or voltage \cite{Vaskivskyi:2016} pulses, and a most probably related ``metallic mosaic'' state created locally by STM pulses \cite{Ma:2016,Cho:2016}.

$1T-\mathrm{TaS_2}$ is a narrow-gap Mott insulator existing unusually on the background formed by a sequence of CDW transitions which have gaped
most of the Fermi surface of the high temperature  metallic (with 1 electron per Ta site) parent phase \cite{Rossnagel:2011}.
Incomplete nesting leaves each 13-th electron ungaped which in a typical CDW would give rise to a pocket of carriers. Here, each excess carrier is self-trapped by inwards displacements of the surrounding atomic hexagon (forming the ``David star'' unit) which gives rise to the intragap local level accommodating this electron.
Exciting the self-trapped electron from the intragap level deprives the deformations from reasons of existence, the David star levels out in favor of a void in the crystal of polarons.
The charged voids are expected to arrange themselves into a Wigner crystal subjected to constraints of commensurability and packing with respect to the underlying structure.

A major question arises: why and how the repulsive voids aggregate into the net of walls leaving micro-crystalline domains in-between?
In this paper, we answer this and related questions by modeling the superlattice of polarons upon the 2D triangular basic lattice of all Ta atoms by a classical charged lattice gas with a screened repulsive Coulomb interaction among the particles.
The external pulse injecting the voids was simulated by introducing a small random concentration of voids reducing the particles concentration $\nu$ below the equilibrium $\nu_0=1/13$ (some other experimentally relevant concentrations are briefly described in the Supplementary Material, Sec. III).
The subsequent evolution of the system, including the passage through the thermodynamic first order phase transition, was studied by means of the Monte Carlo simulation.
Surprisingly, this minimalistic model is already able to capture the formation of domain walls in a close visual resemblance with experimental observations and also to explain the effect qualitatively as an intriguing result of the charge fragmentation.

\section{The model}

We model the system of polarons by a lattice gas of charged particles on a triangular lattice. Each particle represents the self-trapped electron in the middle of the David star, thus the effective charge is $-e$, which is compensated by the static uniform positive background.

The external pulse is simulated by a small concentration of randomly seeded voids reducing the particles' concentration below the equilibrium: $\nu=\nu_0 - \delta \nu$.
The interaction of polarons located at sites $i,j$ is described by an effective Hamiltonian $H = \sum_{i,j} U_{ij} n_i n_j$ with repulsive interactions $U_{ij}$. Here the sum is over all pairs of sites $i \ne j$; $n_i = 1$ (or $0$) when particle is present (or absent) at the site $i$, and we choose $U_{ij}$ as the screened Coulomb potential
%
\begin{align}
U_{ij} = \frac{U_0 a}{|{\mathbf r}_i - {\mathbf r}_j|} \exp\left(-\frac{r-a}{l_s}\right),
\label{potentialU}
\end{align}
%
where $U_0 = e^2 \exp(-a/l_s)/a$ is the Coulomb energy of interaction of particles at neighboring sites in the Wigner crystal state with the distance $a=\sqrt{13}b$ between them ($b$ is the lattice spacing of the underlying triangular lattice, Fig. \ref{fig_lattice_void}a), $l_s$ is the screening length.
We keep in mind also the background uniform neutralizing positive charge.


\section{Superlattice and its charged defects.}

\begin{figure}[tbh]
\centering
\subfloat[]{%
  \includegraphics[width=0.51\linewidth]{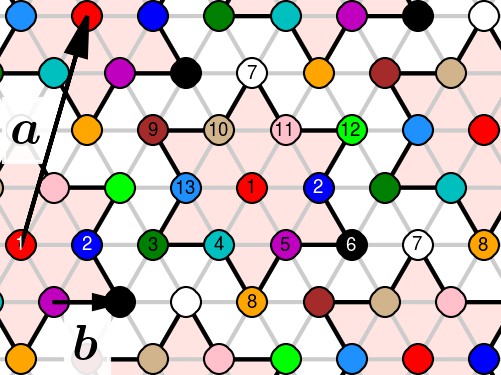}}
\hfill
\subfloat[]{%
  \includegraphics[width=0.43\linewidth]{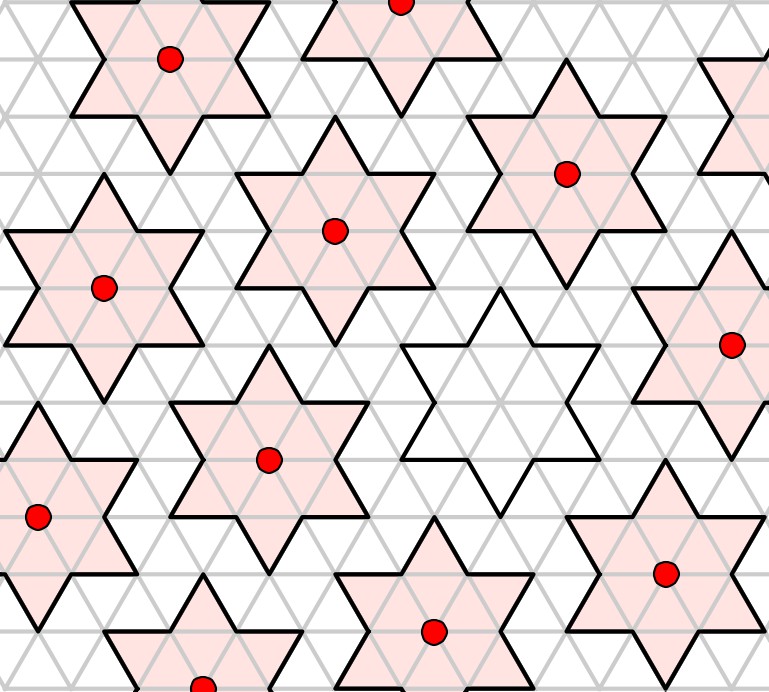}}
\caption{ (a) Underlying triangular lattice (with the period $b$) of all Ta-atoms and the coloring scheme showing its 13 sublattices; in the ground state only one sublattice (with the period $a$) is occupied.
(b) Lattice with $\nu = 1/13$  concentration of polarons at presence of one void (the emptified "David star"). The occupied polaronic sites, marked by red circles, are surrounded by filled "David stars" which perimeters pass through neighbors which positions are actually displaced inwards  (not shown here).}
\label{fig_lattice_void}
\end{figure}

In the ground state, all particles living on the triangular lattice tempt to arrange themselves in also the triangular superlattice, which is close-packed and most energetically favorable in 2D \cite{Bonsall:1977} (with some notable exceptions for more exotic potentials \cite{Jagla:1999}).
Since the concentration of particles is $1/13$, then the ground state is 13-fold degenerate with respect to translations (Fig. \ref{fig_lattice_void}a). An additional mirror symmetry makes the ground state to be in total 26-fold degenerate. But since within a given sample two mirror-symmetric phases do not coexist both in the experiment \cite{Shiba:1986} and in the modeling for the sufficiently slow cooling rates (because of the high energy of the corresponding twinning wall), then we consider only one of them.

\begin{figure}[tbh]
\centering

\subfloat[$q=+e/13$]{%
  \includegraphics[width=0.47\linewidth]{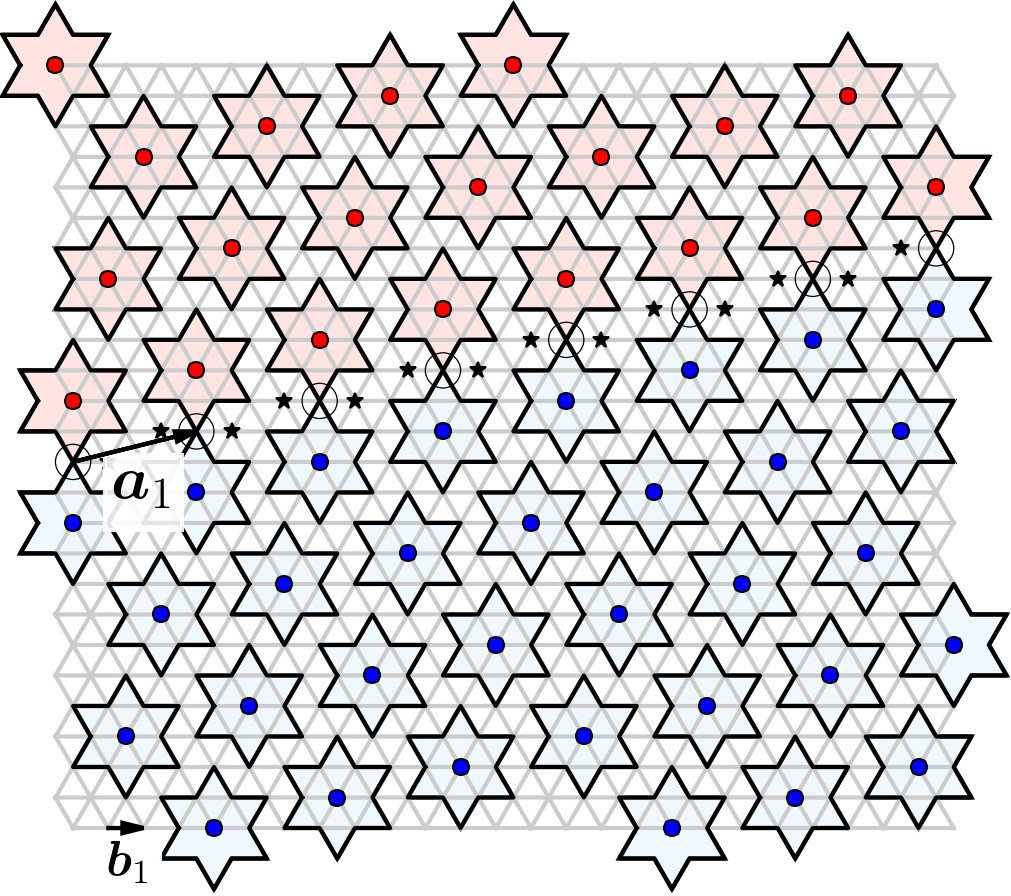}%
}
\hfill
\subfloat[$q=+2e/13$]{%
  \includegraphics[width=0.47\linewidth]{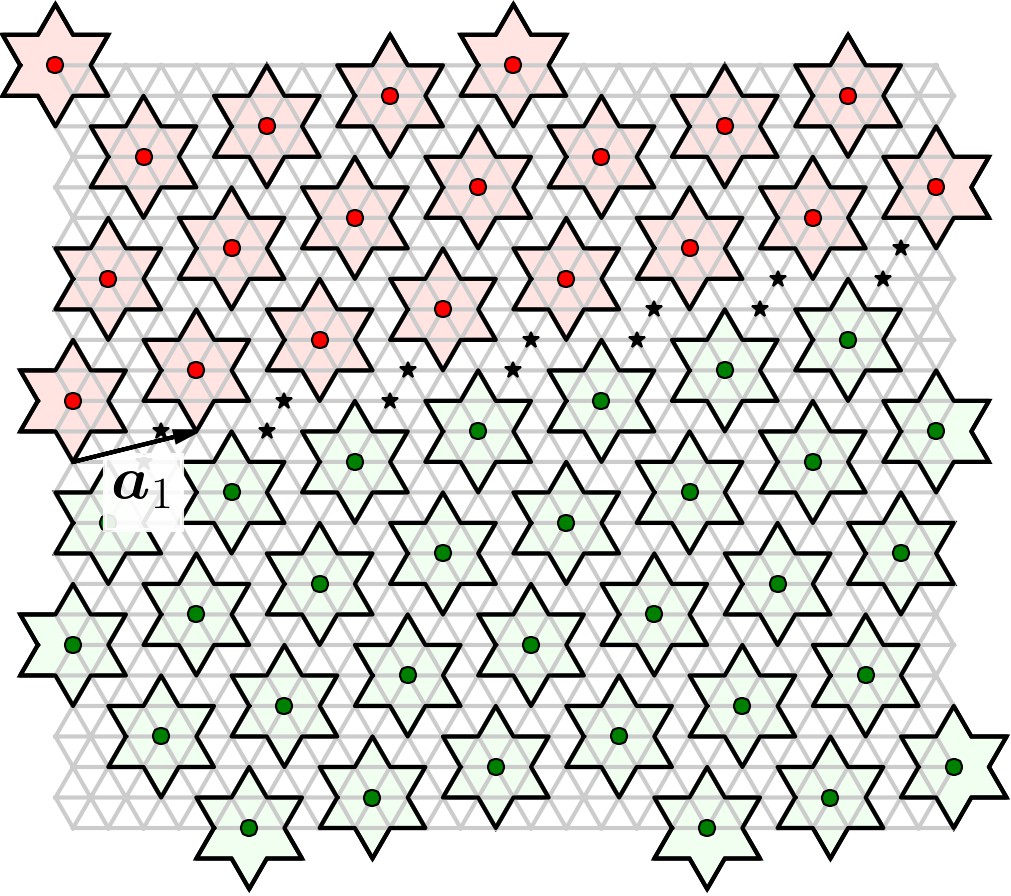}%
}

\caption{Positively charged domain walls with charges per unit cell length ${\mathbf a}$: (a) $+e/13$ ; (b) $+2e/13$. The whole sequence of domain walls  can be obtained by consecutive displacements of the blue domain by the vector ${\mathbf b}_1$ as indicated in (a). In (a), the sites are encircled where David stars within the wall share the corners. Black asterisks designate the sites not belonging to any star.}
\label{fig_wall_charged}
\end{figure}

The simplest lattice defect is a void or a ``polaronic hole'' (Fig. \ref{fig_lattice_void}b) which is formed when the electron from the intragap level is taken away or excited to the conduction band and soon the associated lattice distortions vanish. The single void has the relative charge $+e$  (keeping in mind the background neutralizing charge) and the Coulomb self-energy of the order
$E_{void} \simeq e^2/a$.
While the void is a particular manifestation of a general notion of vacancies in crystals, in our case there can be also a specific topologically nontrivial defect -- the domain wall separating domains with a different 13-fold positional degeneracy of the ground state (Fig. \ref{fig_wall_charged}).
The domain wall cross-section resembles the discommensuration known in CDW systems \cite{McMillan:1976}.

Experimentally, the lattice defects can be introduced via external pulses, by impurity doping or by the field effect. For example, a laser or STM pulse can excite the Mott-band electrons residing in the centers of the David star clusters, creating an ensemble of voids. Since the voids are charged objects, then at first sight they should repel each other and form a Wigner crystal themselves.
But our modeling consistent with the experiment shows that the voids rather attract one another at short distances and their ensemble is unstable towards formation of domain walls' net.
Qualitatively, this instability can be understood from the following argument.
Compare energies of the isolated void and of the domain wall segment carrying the same charge.
The minimal charge of domain wall per the translation vector ${\mathbf a_1}$ is $+e/13$ (Fig. \ref{fig_wall_charged}a), and the energy of the wall's segment carrying the charge $+e$ can be estimated as for a uniformly charged line:
%
\begin{align}
E_{wall} \simeq 13 \times \frac{(e/13)^2}{a} \ln (l_s/a),
\label{Ewall}
\end{align}
%
For moderate screening lengthes $l_s$, it is lower than the void's self-energy $E_{void}$, making energetically favorable to decompose the voids into fractionally-charged domain walls. The local effects beyond our model can also favor domain walls with other charges: for the single-step $+1e/13$ domain wall there are anomalous sites where David stars intersect (Fig. \ref{fig_wall_charged}a) which raises its energy and can make the double-step $+2e/13$ domain walls (Fig. \ref{fig_wall_charged}b) to be energetically favorable .

%
%


\section{Numerical modeling}

We simulated the cooling evolution of the classical lattice gas with the interaction potential (\ref{potentialU}) via Metropolis Monte Carlo algorithm (see Methods).
We perform slow cooling from $T=\Tstart{}$, which is above the detected ordering phase transition (see below), down to $T=\Tstop{}$ with a step $\Delta T=\dT{}$, reaching either a ground state or a very close in energy metastable state.
Below we, first, consider undoped systems (where particles concentration $\nu$ is exactly $\nu_0=1/13$), and then systems doped by voids (with $\nu=\nu_0-\delta\nu \equiv \nu_0 (1-\nu_{voids})$, where $\nu_{voids}=\delta\nu/\nu_0$ is the voids' concentration). Results for another sign of doping are briefly presented in the Supplementary Material, Sec. III.



\subsection{Undoped system.}

As a reference system we chose the sample with $91\times104$ sites with the total a number of particles $N_{p}=728$, which corresponds to the concentration $\nu_0=1/13$.

\begin{figure}[tbh]
\centering
\subfloat[]{
  \includegraphics[width=0.49\linewidth]{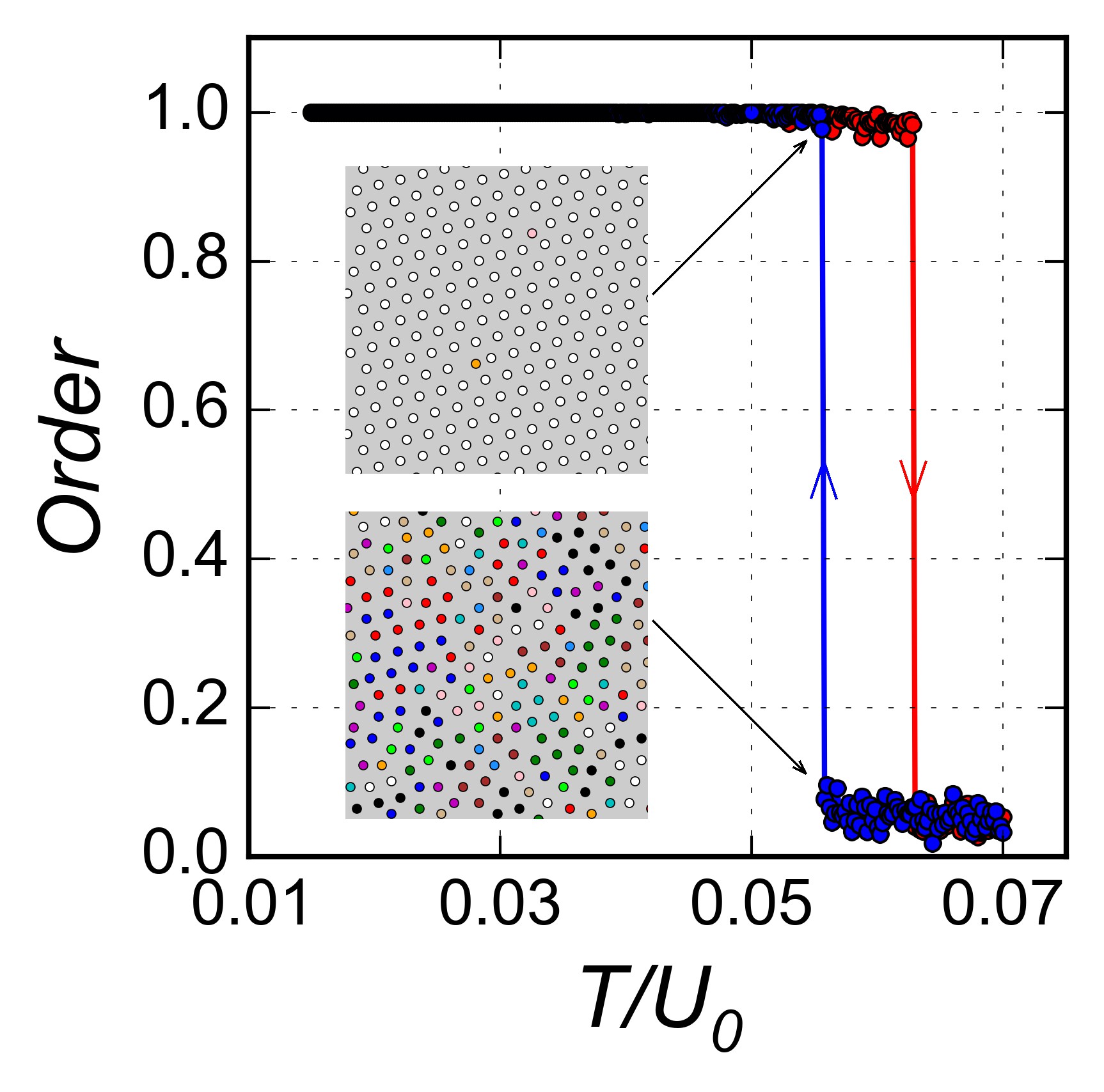}
}
\subfloat[]{
  \includegraphics[width=0.49\linewidth]{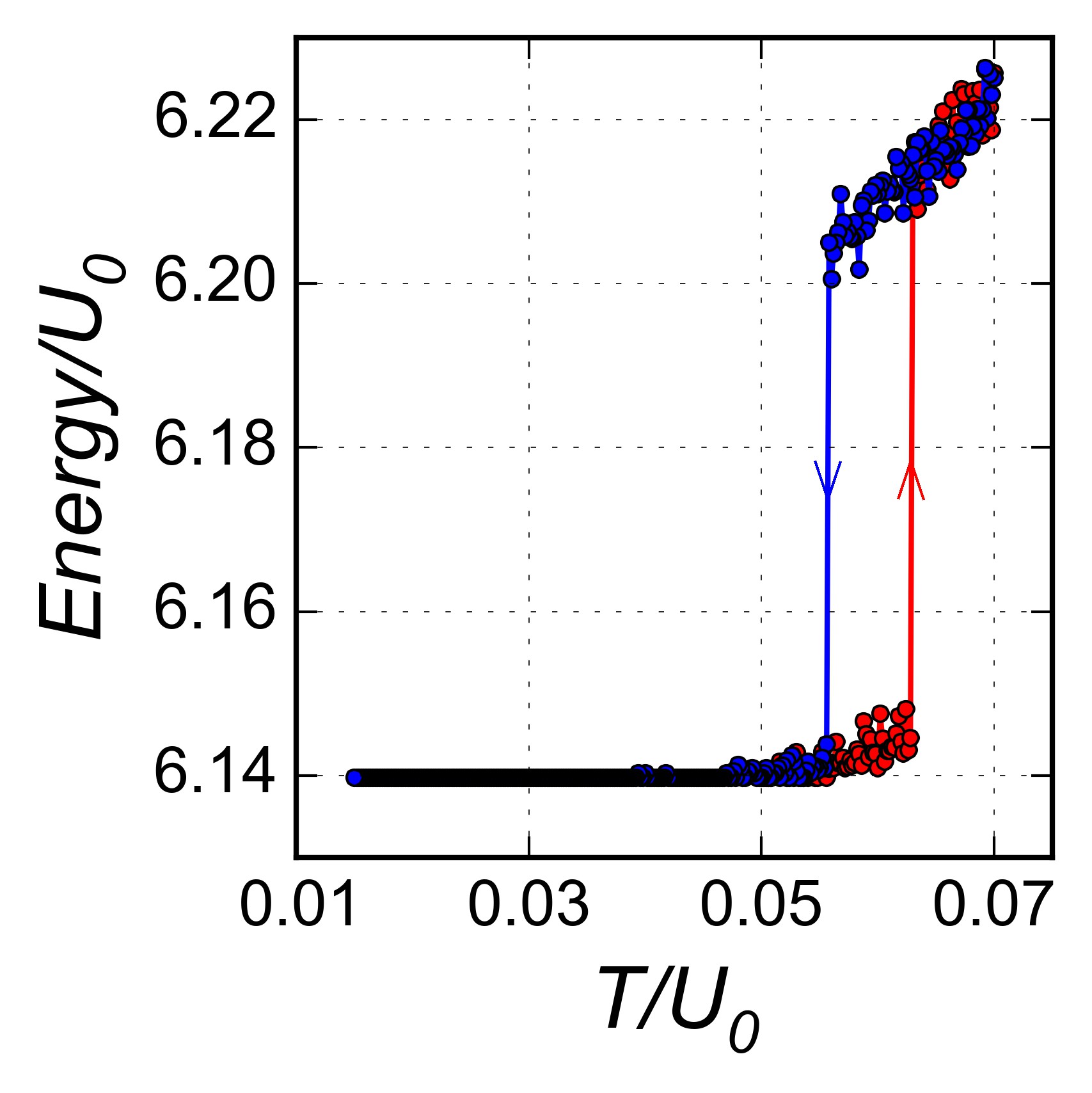}
}

\caption{Temperature dependencies of integrated characteristics for the undoped system. (a) The order parameter; the insets show snapshots of configurations of the system just above and below the phase transition; (b) mean energy per particle. Blue symbols are for cooling and red symbols are for heating simulations.}
\label{M_E_vs_T}
\end{figure}

On cooling, the order-disorder phase transition takes place at $T_c \approx \Tc{}$, below which the triangular superlattice is formed confirming the expectations for the Wigner crystal. Temperature dependencies of the order parameter
$M = \sqrt{\sum(m_i-1/13)^2/13 \cdot 12}$, where $m_i$ is the fraction of particles at $i$-th sublattice (Fig. \ref{M_E_vs_T}a) and of the mean value of energy per particle (Fig. \ref{M_E_vs_T}b) indicate that the transition is of the first order. The insets in the Fig. \ref{M_E_vs_T}a show a plenty of defects just above $T_c$, while only two displaced positions are left just below $T_c$.

On heating, the order-disorder phase transition takes place at $T\approx0.063 U_0$, which agrees with our mean field analysis (see Supplementary Sec. II).
With increasing $l_s$ the temperature hysteresis and the tendency to overcooling become more pronounced.
An overcooling or even freezing into in a glass state feature is known for electronic systems with either a frozen disorder or a Coulomb frustrations \cite{Dobrosavljevic:2015}; however in the present model these both factors are absent -- the effect is presumably due to only the long-range Coulomb interactions under the lattice constraints.

\subsection{Doped system.}
We emulate the doping (the charge injection) by seeding voids at random places and following the subsequent evolution. By global characterizations like those in Fig. \ref{M_E_vs_T}, the order-disorder transition is preserved, while at a lower temperature. But locally the new mosaic ground state with the net of domain walls is formed as we will demonstrate below.

Seeding at $T<T_c$ a small number of voids, down to two defects per sample, we observe that the single-void states are unstable with respect to their binding and progressive aggregation.
Seeding more voids initiates their gradual coalescence into a globule of interconnected segments of domain walls .
The resulting globule performs slowly a random diffusion over the sample while keeping closely its optimal shape and the structure of connections (the Supplementary Video of the system evolution under cooling and its description are presented in Supplementary Sec. I).

Figure \ref{fig_globule}a shows the low-$T$ configuration of system $130\times156$ system, where initially $N_p=1544$ of particles (with the the corresponding concentration of voids $\nu_{voids}\approx\dopingGlobulePercent{}\%$) were randomly seeded and then the system was slowly cooled from $T=\Tstart{} > T_c$ down to $T=\Tstop{}$. In spite of the initial random distribution of particles over the whole sample, finally the voids aggregate into a single globule immersed into a connected volume of the unperturbed crystal.
We compare the results of our modeling in Fig. \ref{fig_globule}a with the experimental picture in Fig. \ref{fig_globule}b  \cite{Ma:2016}. Similar patterns have been observed also in other experiments: \cite{Vaskivskyi:2016} (Fig. 3 of the supplement) and \cite{Cho:2016} (Fig. 1).

With a further increase of doping, the globule size grows over the the whole sample, and finally the branched net of domain walls divides the system into the mosaics of randomly shaped domains (Fig. \ref{fig_net}a,c).
The comparison of the modeling with the experiment on injection by the STM pulses is shown between panels (a) and (b) in Fig. \ref{fig_globule}, between panels (a) and (b), (c) and (d) in Fig. \ref{fig_net}. The figures visualize a spectacular resemblance of our modeling with results from several STM experiments exploiting either the optical switching to the hidden state \cite{Gerasimenko:2017} or the pulses from the STM tip \cite{Ma:2016,Cho:2016}.
Note that similar ``irregular honeycomb network'' structures were predicted for incommensurate phase of krypton on graphite with $\nu_0\approx1/3$ and short-range interaction \cite{Villain:1980}, but with less topological restrictions here.

\begin{figure}[tbh]
\centering
\subfloat[]{\includegraphics[width=0.51\linewidth]{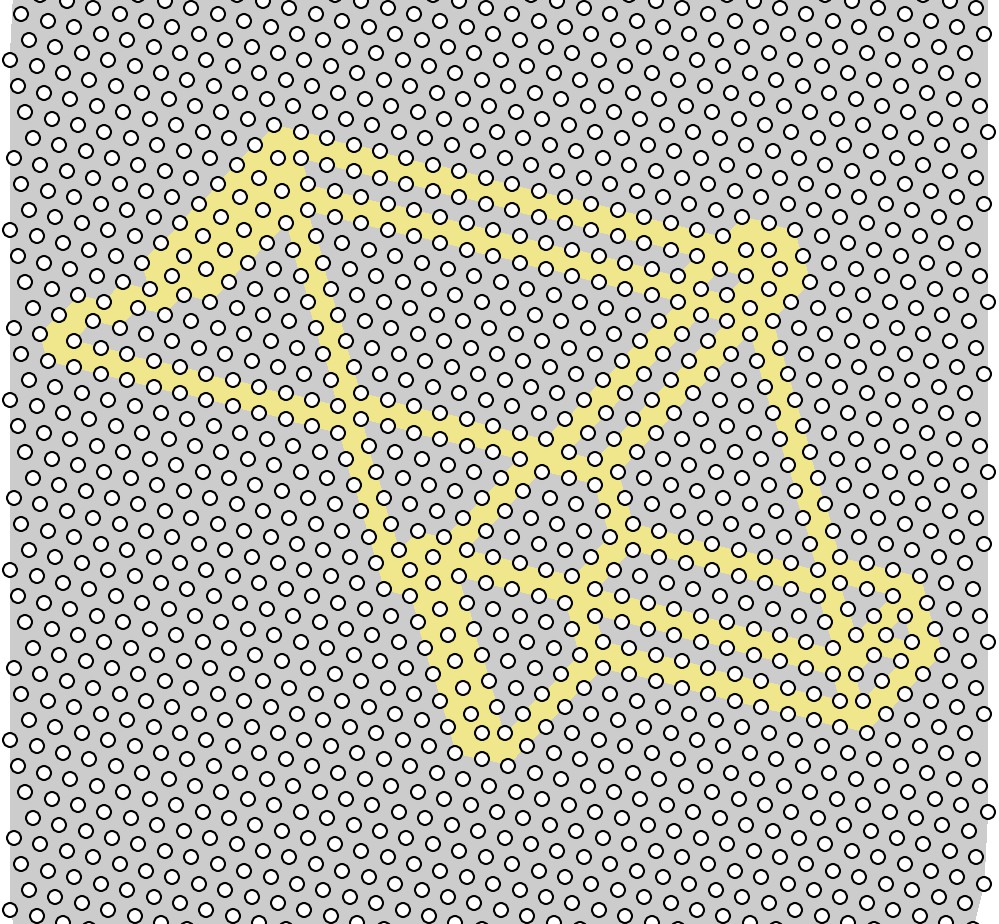}}
\hfill
\subfloat[]{\includegraphics[width=0.48\linewidth]{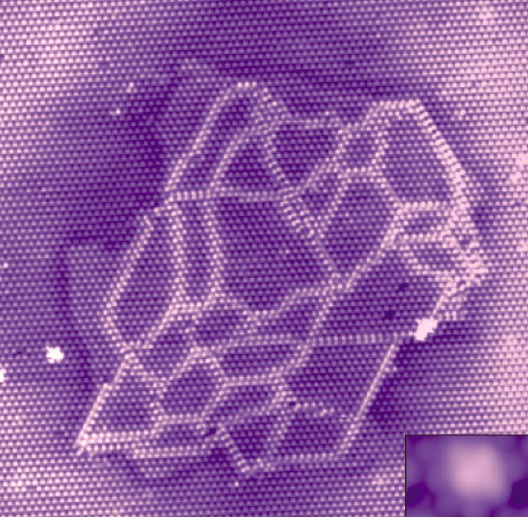}}
\caption{Globule structures. (a) The present modeling for $\nu_{voids} \approx \dopingGlobulePercent{} \%$
in the domain walls representation, $T=0.01 U_0$; (b) from experiments in \cite{Ma:2016}.}
\label{fig_globule}
\end{figure}

\begin{figure}[tbh]
\centering
\subfloat[$\nu_{voids} \approx \dopingNetPercent{} \%$ -- domain walls]{\includegraphics[width=0.48\linewidth]{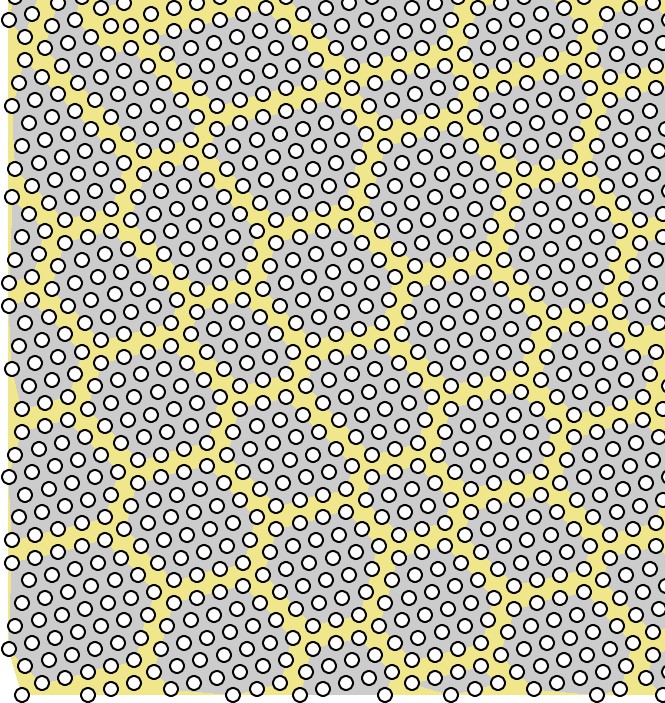}}
\hfill
\subfloat[]{\includegraphics[width=0.48\linewidth]{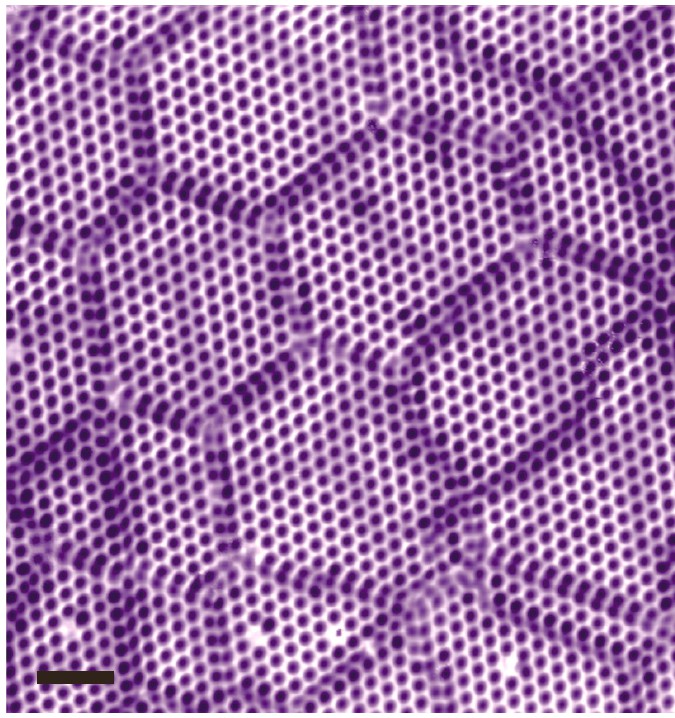}}

\subfloat[$\nu_{voids} \approx \dopingNetPercent{} \%$ -- domains]{\includegraphics[width=0.48\linewidth]{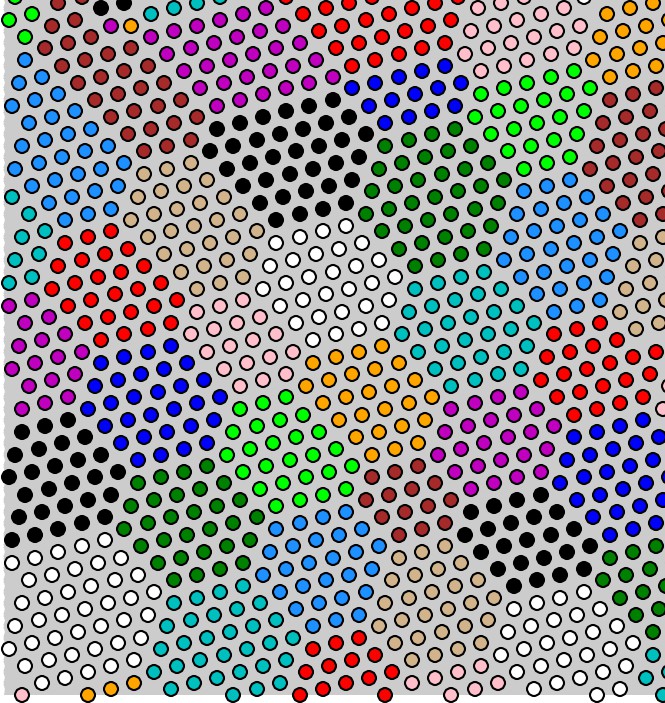}}
\hfill
\subfloat[]{\includegraphics[width=0.47\linewidth]{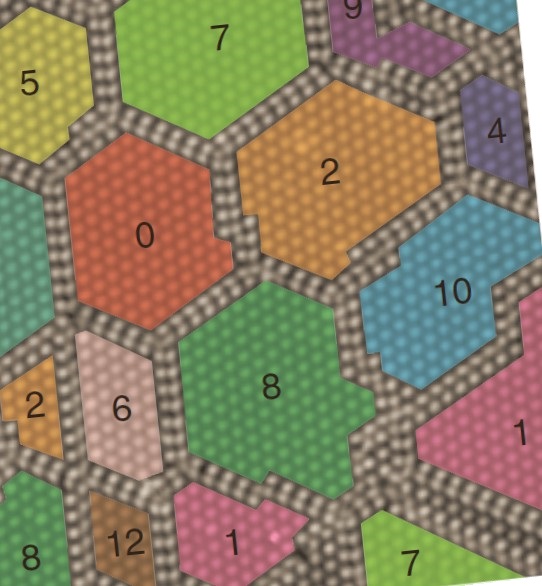}}

\caption{The modeling for a high doping (a,c) vs experiments (b,d). Maps of domain walls (a,b) and of domains (c,d).
Figs. (a,c) show the present modeling with $\nu_{voids}=\dopingNetPercent{}\%$ of voids at low temperature $T=0.01 U_0$, the coloring scheme for domains is indicated in Fig. \ref{fig_lattice_void}.
Figs. (b,d) are adapted from \cite{Ma:2016}, numbers in Fig. (d) show the corresponding coloring scheme for domains.
Figs.(a,c) show only a part of the system, full images can be found in Supplementary Fig. 6.}
\label{fig_net}
\end{figure}

\section{Discussion and Conclusions}

Our simulations have shown an apparently surprising behavior: some effective attraction of voids develops from the purely repulsive Coulomb interactions.
The coalescence of single voids starts already at their small concentration. For several voids seeded, we observe a gradual fusion of point defects into the globule of the domain walls.
Increasingly branched net develops with augmenting of the voids concentration.

That can be understood indeed by noticing that the walls formation is not just gluing of voids but their fractionalization. The domain wall is fractionally ($q=\nu_0 e$) charged per its crystal-unit length, thus reducing the Coulomb self-energy in comparison with the integer-charged single void. Being the charged objects, the domain walls repel each other, but as topological objects they can terminate only at branching points, thus forming in-plane globules. Their repulsion at adjacent layers meets no constraints, hence the experimentally observed alternation of the walls' patterns among the neighboring layers \cite{Gerasimenko:2017, Ma:2016, Cho:2016}.

A similar while simpler doping induced phase transition to the state patterned by charged domain walls was predicted for for quasi-1D polyacethylene-like systems with 2-fold degeneracy, see \cite{Karpov:2016}.

Our modeling can be straightforwardly extended to other values of concentration $\nu_0$: like $\nu_0=1/3$ which is the minimal value where the pattern formation appears with qualitatively similar to the presented here results (see Supplementary Sec. III) and corresponding, for example, to the $\sqrt{3}\times\sqrt{3}$ surface CDW observed in the lead coated germanium crystal \cite{Carpinelli:1996}.
For another experimentally known case of $2H-\mathrm{TaSe_2}$ where $\nu_0=1/9$ the modeling results are quite different: in some doping range we see a ``stripe'' phase (see Supplementary Sec. III), which indeed was experimentally observed in this material \cite{Fleming:1980}. The exception of the case $\nu_0=1/9$ is rather natural, because here the basis vectors of the superlattice and of the underlying triangular lattice are parallel to each other, which allows for the existence of neutral elementary domain walls.

It is also possible to study the doping by electrons by seeding the interstitials rather than voids; here the new David stars substantially overlap with their neighbors giving rise to stronger lattice deformations, which may require for a more complicated model. The fragmentation with formation of walls is always confirmed while details of patterns can differ (see Supplementary Fig. 7).

The encouraging visual correspondence of our pictures with experimentally obtained patterns in different regimes of concentrations ensures a dominant role of the universal model.
\\

\textbf{Methods}

For numerical simulation of the classical lattice gas with interaction potential (\ref{potentialU}) we employed the Metropolis Monte Carlo method. We used the screening parameter $l_s = 4.5 b \approx 1.25 a$ and truncated the interactions to zero for sufficiently large interparticle distances (outside the hexagon with the side $24 b$).
At each temperature we performed $\sim 10-40$ millions of Monte Carlo steps depending on the numerical experiment. 
Temperature was linearly lowered from $T=\Tstart{}$ down to $T=\Tstop{}$ with a step $\Delta T=\dT{}$.
The following system sizes and numbers of particles were chosen:
size $91\times104$ and $N_p=728$ for undoped system;
size $130 \times 156$  and $N_p=1544$, $\nu_{voids} \approx \dopingGlobulePercent{}\%$ for the globule system (Fig. \ref{fig_globule}a);
size $142\times164$  and $N_p=1758$, $\nu_{voids} \approx \dopingNetPercent{}\%$ for the net system (Fig. \ref{fig_net}a,c).
Periodic boundary conditions were imposed.
\\

\textbf{Acknowledgements}

The authors are grateful to D. Mihailovich, Y.A. Gerasimenko,  E. Tosatti and H.W. Yeom for helpful discussions. We acknowledge the financial support of the Ministry of Education and Science of the Russian Federation in the framework of Increase Competitiveness Program of NUST MISiS (N K3-2017-033). SB acknowledges funding from the ERC AdG ``Trajectory''.
Image source for Figs. \ref{fig_globule}(b), \ref{fig_net}(b),(d) is \cite{Ma:2016}; use permitted under the \href{https://creativecommons.org/licenses/by/4.0/}{Creative Commons Attribution License CC BY 4.0}; the images were cropped and rotated.
\\

\textbf{Author contributions}

S.B. and P.K. together formulated the theoretical concept, designed the model, analyzed results, and wrote the paper; P.K. performed the numerical computations. Correspondence and requests for materials should be addressed to P.K.
\\

\textbf{Additional information}

The authors declare no competing financial interests.

\end{document}


\title{Supplemental Material. \\
Modeling of networks and globules of charged domain walls observed in pump and pulse induced states.} 

\author{Petr Karpov}
\email{karpov.petr@gmail.com}
\address{National University of Science and Technology ``MISiS'', Moscow, Russia}
\author{Serguei Brazovskii}
\address{National University of Science and Technology ``MISiS'', Moscow, Russia}
\address{CNRS UMR 8626 LPTMS, University of Paris-Sud, University of Paris-Saclay, Orsay, France}
\address{Jozef Stefan Institute, Jamova 39, SI-1000 Ljubljana, Slovenia}

\date{November 2, 2017}

\maketitle

\section{Movie: coalescence of voids and cluster diffusion}

\begin{figure}[tbh]
\centering
\subfloat[]{%
  \includegraphics[width=0.49\linewidth]{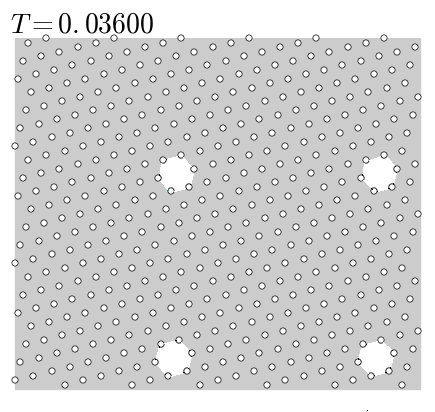}%
}
\hfill
\subfloat[]{%
  \includegraphics[width=0.49\linewidth]{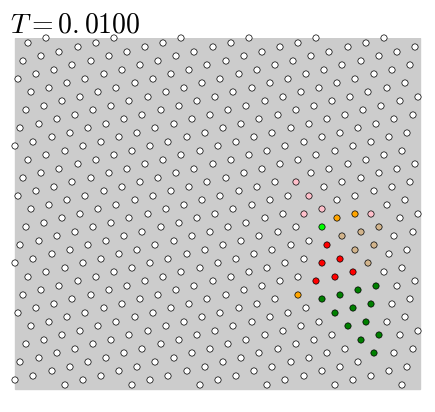}%
}
\caption{(a) The initial position of the simulation at $T=0.036 U_0$, holes' locations are highlighted by white background; (b) the final position of the simulation at $T=0.010 U_0$.}
\label{fig_coalescence}
\end{figure}

The movie (\underline{\href{https://www.dropbox.com/s/wv1ebnjys3bm0da/suppl-movie.mp4}{https://www.dropbox.com/s/wv1ebnjys3bm0da/suppl-movie.mp4}}) reflects the process of Monte Carlo (MC) cooling of the system, after several voids have been added to the ordered state below the transition temperature.
This procedure differs from the one exploited to obtain the final states reported in the main text. There we always started from a random disordered state at $T > T_c$ in order to reach a better equilibration.
The movie demonstrates two regimes in the course of the cooling: (1) merging of voids into a single big cluster seen as the globule of domain walls; (2) random diffusion of the cluster.

We start the simulation from the ordered system where 4 voids have been seeded by hands. (Fig. \ref{fig_coalescence}a shows the initial position) and cool it from $T=0.036 U_0$ down to $T=0.010 U_0$ with step $\Delta T=-0.0001 U_0$,  making at each temperature 10000 MC steps for higher temperatures ($T \geq 0.0310 U_0$) or 50000 MC steps for lower temperatures ($T < 0.0310 U_0$).

First, each of 4 voids very quickly disintegrates into a small cluster (``1-cluster''), and then begins a random diffusion (note, that periodical boundary conditions are imposed).
Being the charged objects, the 1-clusters are repelled from each other.
Nevertheless, if two 1-clusters overcome a potential barrier, they merge  into a stable 2-cluster, this which lowers the system energy (the qualitative explanation of this counterintuitive effect is given in the main body of the paper). The first event of two 1-clusters merging happens at $T=0.03548 U_0$ after $\sim 52 000$ MC steps (5-th second of the video).

After several collisions and unsuccessful coalescence attempts, the 2-cluster merges with a 1-cluster at $T = 0.03407 U_0$ (after $\sim 193 000$ MC steps, 19-th second of the video). Final merging to a single cluster happens shortly after that at $T=0.03378 U_0$.

Afterwards, the resultant big cluster performs a random diffusion through the sample. After the 50-th second, the video is speeded up by the factor of 50. Figure \ref{fig_coalescence}b shows the final position of the simulation.

We conclude that it is indeed energetically favourable and dynamically sustainable for separate holes to merge into a single cluster.

\section{Order-disorder phase transition: mean field theory}

In this section we present the mean field theory for the order-disorder phase transition for an undoped system.
Since the interactions are of the long-range nature and many particles affect the given one, we expect mean-filed theory to be a good approximation.

At $T=0$, perfectly ordered state is observed, where all particles occupy the same sublattice (one of the 13 equivalent sublattices, not counting for the mirror symmetry, as discussed in the main text). With gradually increasing the temperature,
at some critical temperature $T=T_c$ order momentarily breaks.

\begin{figure}[tbh]
\centering
  \includegraphics[width=0.35\linewidth]{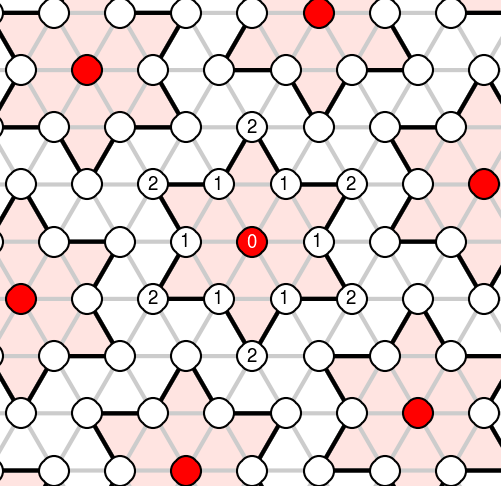}%
\caption{One of the ground states of the system, where particles (red circles) occupy the same sublattice. 1-st, and 2-nd coordination spheres for a given particle (``0'') are shown.}
\label{fig_coordination}
\end{figure}

In order to determine  $T_c$ we consider a simplified version of the model, dividing the whole system into closely packed David stars, and allowing for each particle to occupy only one of 13 sites of its David star (Fig. \ref{fig_coordination}). This simplification will lead to a slight overestimation for $T_c$. 

Since the energy $\Delta_1$ (``gap'') to excite a particle from its regular position
in the David star center to the first coordination sphere is several times less than the analogous energy for the second coordination sphere $\Delta_2$ (for example, for the studied in the main text case $l_s = 4.5 b$ we have found $\Delta_1 \approx 0.442 U_0$, $\Delta_2 \approx 1.993 U_0$). Then taking into account only 1-st coordination sphere seems to be a reasonable first approximation and we arrive at a simpler model with only 7 possible states for each particle.

Consider a particle in the mean field of other particles. Let $(s_0, s_1)$ be the occupation numbers of the 0-th and the 1-st coordination spheres respectively. In the ground state we have $s_0 = 1$, $s_1 = 0$; in the excited state $s_0 = 0$, $s_1 = 1$. Introduce the order parameter
\begin{equation}
m = \frac{7}{6} \langle s_0 - \frac{1}{7} \rangle,
\label{self_consistency}
\end{equation}
so in the ordered phase $m=1$, in the disordered phase $m=0$.
Since on average only those particles that occupy the given sublattice try to keep the particle in the same sublattice, we use the following ``temperature-dependent gap''
%
$\Delta(T) = \Delta_1 \cdot m(T).$
%
The one-particle mean-field Hamiltonian
$H = s_1 \Delta_1 m(T)$,
gives rise to the partition function  $Z(T) = 1 + 6 e^{-\Delta_1 m/T}$.
From the self-consistency condition (\ref{self_consistency}), we get
$m =  \left( 1 -  \frac{7}{6} e^{-\Delta_1 m/T}  \right) /Z$ or
\begin{equation}
m = \frac{1-e^{- m \Delta_1/T }}{1+6 e^{- m \Delta_1/T }}
\label{self_consistency2}
\end{equation}
Expanding RHS of (\ref{self_consistency2}) to the first order in $m$ we get $m = m \Delta_1/7 T + o(m)$,
which yields to the critical temperature
\begin{equation}
T_c = \frac{\Delta_1}{7}.
\label{self_consistency3}
\end{equation}
For $l_s = 4.5 b$ we have $\Delta_1 \approx 0.442 U_0$, and formula (\ref{self_consistency3}) gives us $T_c \approx 0.0631 U_0$, which is very successfully compared with the Monte Carlo simulation result on heating: $T_c \approx 0.0630 U_0$.

We conclude that the mean field theory is in a very good agreement (within 0.5\%) with the results of the simulation. This is rather natural because for the screened Coulomb long-range interaction many coordination spheres of particle's neighbors contribute to the mean field for a given particle (this fact is incorporated into the definition of the excitation energy $\Delta_1$).







\section{Results for different values of concentration $\nu_0$}

\begin{figure}[H]
\centering
  \includegraphics[width=0.5\linewidth]{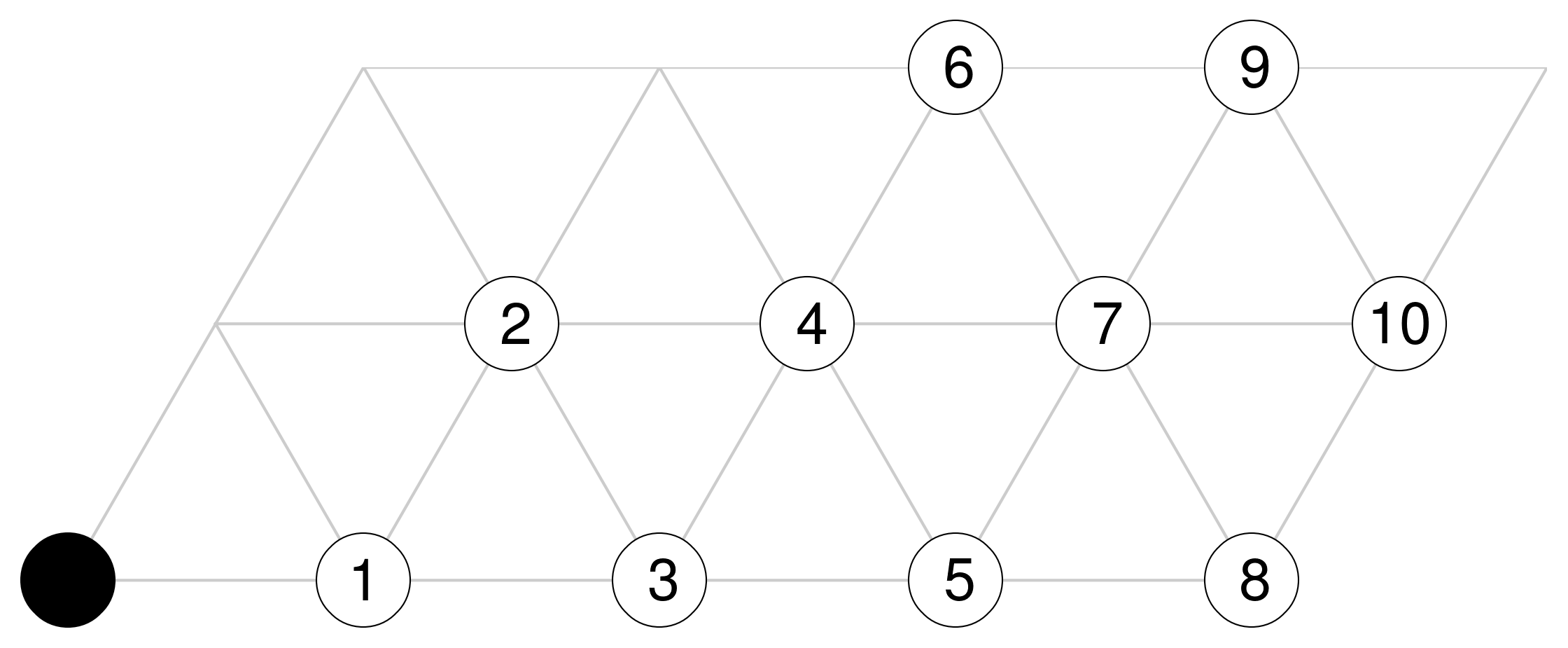}%
\caption{Neighbors of a given site (black) with successively increasing distances from it.}
\label{fig_nearest_neighbors}
\end{figure}

First we give the classification of filling factors on a triangular lattice, and then analyze some representative cases.
Figure \ref{fig_nearest_neighbors} shows ordinal numbers of neighbors for a given particle (black) -- 1 corresponds to the nearest neighbor, 2 to the next-nearest neighbor etc.
Knowledge of where the particle's nearest neighbor sits determines the whole superstructure. Table I presents the correspondence between the number of the occupied sublattice and the concentration of the particles. The third column of the table shows, whether the superlattice possesses only charged domain walls or not: this is connected to the fact, whether the basis vectors of the superlattice are parallel to the basis vectors of the underlying triangular lattice. This is the key factor, determining the domain walls' patterns in doped systems. The forth column shows, whether the superlattice possesses two different sectors of ground states, which can not be obtained one from another by only a translation -- in this case the mirror symmetry is also needed.

In the further subsections we briefly discuss several experimentally accessible cases, with the concentrations $\nu_0=1/3$, $1/9$, or $1/13$.
\begin{table}[H]
\caption{Classification of commensurate filling factors on a triangular lattice} 
\centering 
\begin{tabular}{c c c c} 
\hline\hline 
\# of neighbor & concentration $\nu_0$ &  only charged domain walls? & two mirror symmetric sectors?\\ [0.5ex] 
\hline 
2 & 1/3  & yes & no  \\ 
3 & 1/4  & no  & no  \\
4 & 1/7  & yes & yes \\
5 & 1/9  & no  & no  \\
6 & 1/12 & yes & no  \\ 
7 & 1/13 & yes & yes \\
8 & 1/16 & no  & no  \\
9 & 1/19 & yes & yes  \\
10 & 1/21 & yes  & yes  \\
[1ex] 
\hline 
\end{tabular}
\label{table1} 
\end{table}

\subsection{$\nu_0 = 1/3$}

Concentration $\nu_0 = 1/3$ is the minimal case, where we observe qualitatively similar patterns as for $\nu_0 = 1/13$.
There is, however, an important distinction: single holes are much more stable here and some threshold holes concentration is necessary for the first globule creation.
This case corresponds, for example, to Pb on Ge surface charge density wave \cite{Carpinelli:1996}.

\begin{figure}[H]
\centering
\subfloat[]{\includegraphics[width=0.49\linewidth]{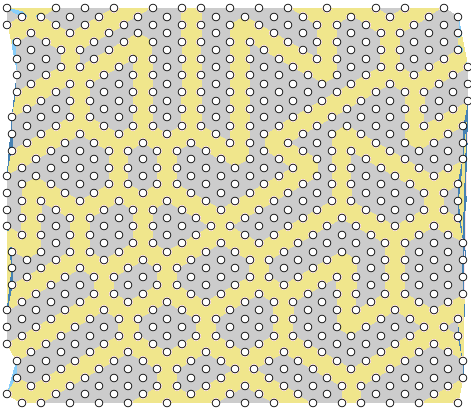}}
\hfill
\subfloat[]{\includegraphics[width=0.49\linewidth]{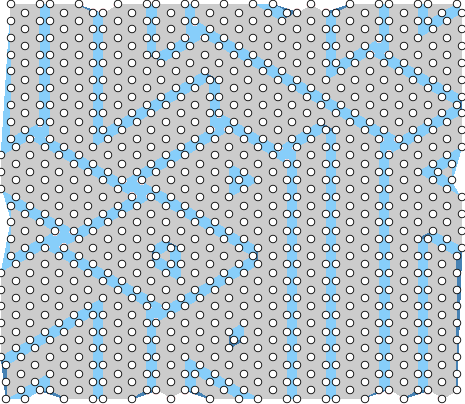}}
\caption{$\nu_0=1/3$: domain wall representation for (a) positive doping; (b) negative doping.}
\label{fig_net_3}
\end{figure}

\subsection{$\nu_0 = 1/9$}

In the case $\nu_0 = 1/9$ the minimal domain walls are neutral, because unit vectors of the superlattice and of the underlying triangular lattice are parallel to each other for this concentration, so the results are qualitatively different from $1/3$ and $1/13$ cases.
Which connectivity is more favorable (stripe or network) depends on the sign of wall-crossing energy \cite{Bak:1982}.
Interestingly the ``stripe'' phase is experimentally observed on heating of $\nu_0 = 1/9$ commensurate charge density wave in $2H-\mathrm{TaSe_2}$, see \cite{Fleming:1980}.

\begin{figure}[tbh]
\centering
  \includegraphics[width=0.4\linewidth]{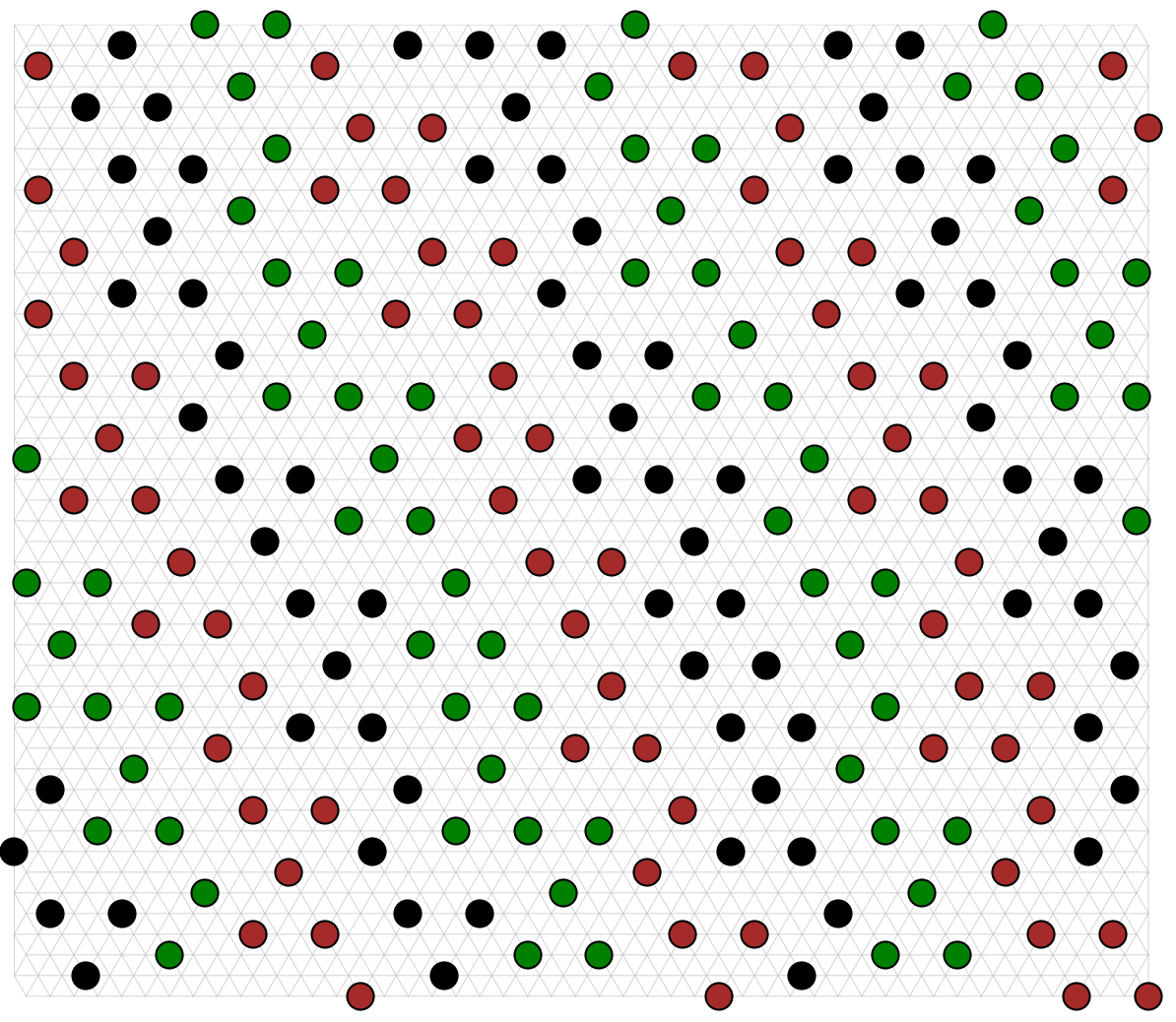}%
\caption{$\nu_0=1/9$: domain representation for the ``stripe'' phase for with positive doping. Here we use 9-coloring scheme (only 3 of them appear on the figure), analogous to 13-coloring scheme used in the main text.}
\label{fig_stripes_9}
\end{figure}

\subsection{$\nu_0 = 1/13$}

\begin{figure}[tbh]
\centering
\subfloat[]{\includegraphics[width=0.48\linewidth]{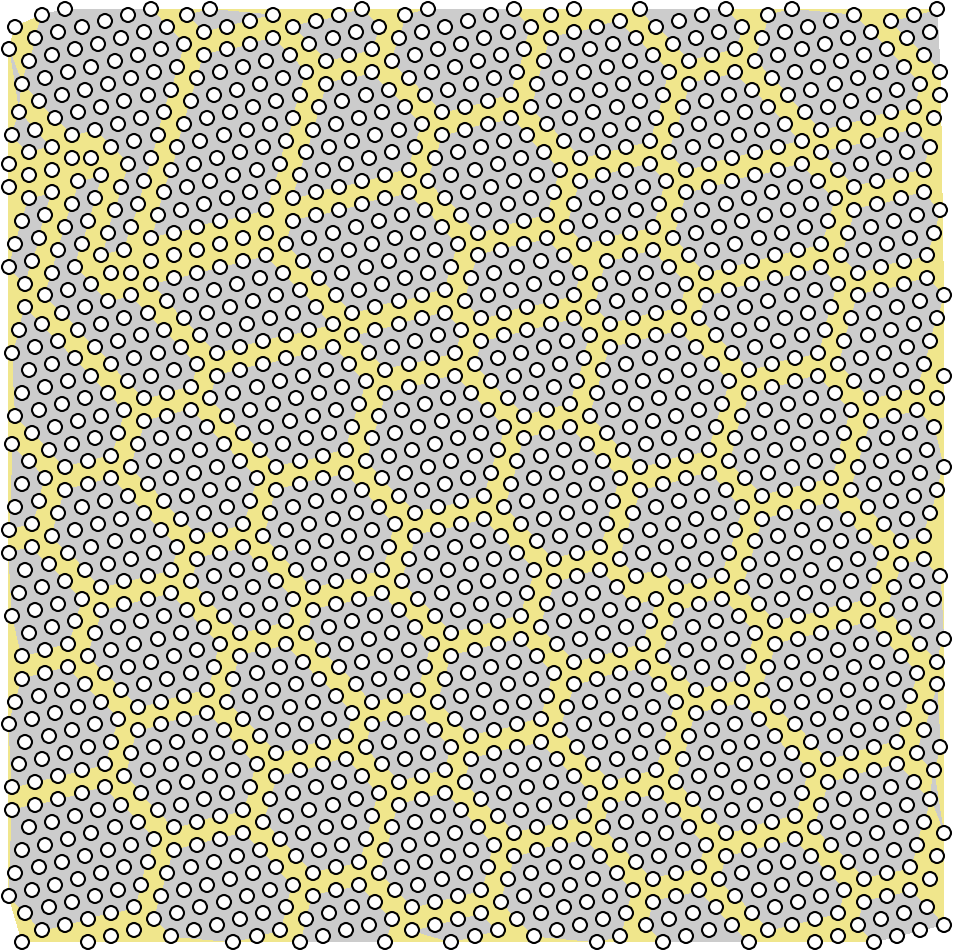}}
\hfill
\subfloat[]{\includegraphics[width=0.48\linewidth]{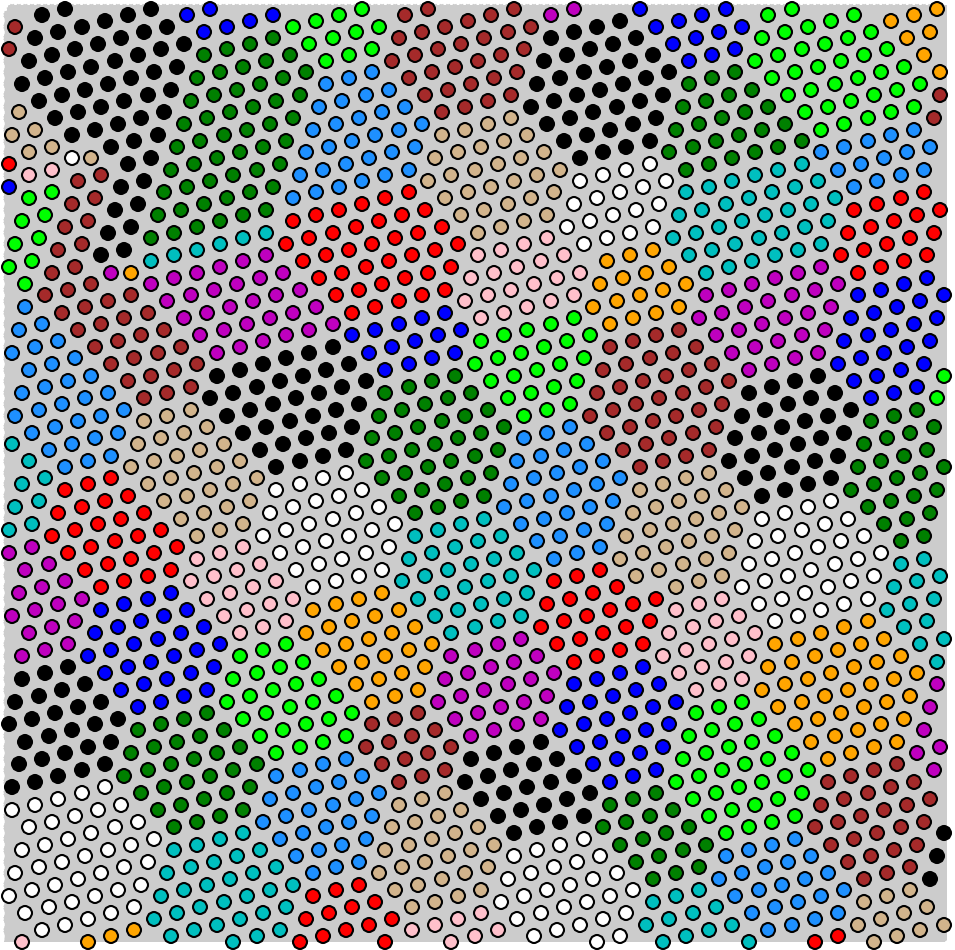}}
\caption{$\nu_0=1/13$: net structure for positive doping: full-sized version of Figure 5 from the main text. (a) Domain wall representation; (b) domain representation}
\label{fig_net_13}
\end{figure}

The case of positive doping was considered in the main text; Fig. \ref{fig_net_13}a,b shows the full-size version of Fig. 5a,c of the main text.

\begin{figure}[tbh]
\centering
\subfloat[]{\includegraphics[width=0.48\linewidth]{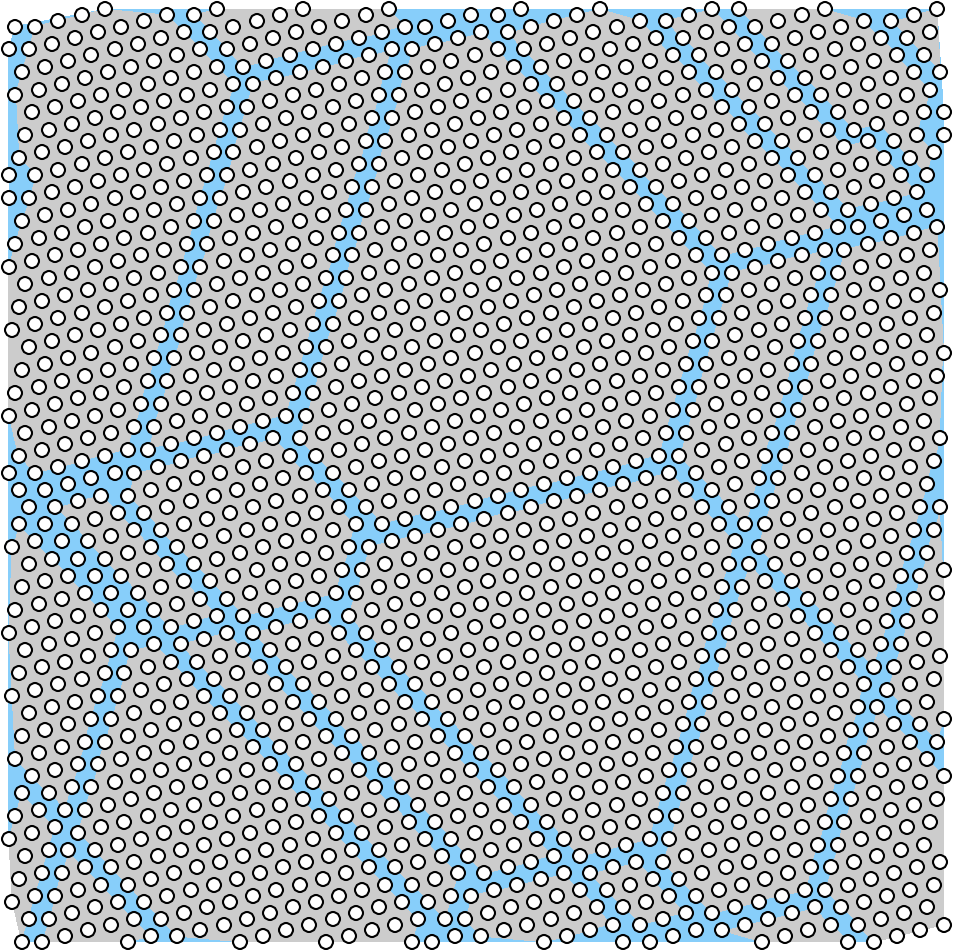}}
\hfill
\subfloat[]{\includegraphics[width=0.48\linewidth]{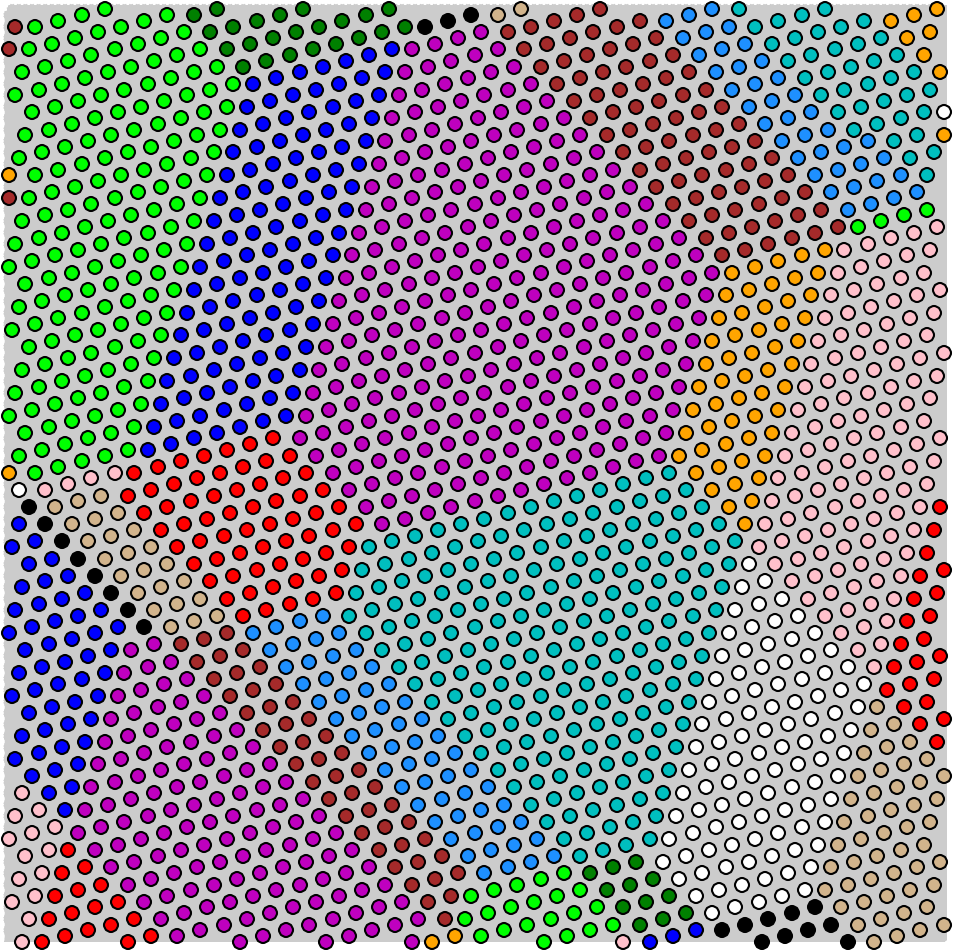}}
\caption{$\nu_0=1/13$: net structure for negative doping, $\nu_{interst} \approx 1.8\%$, $l_s=4.5 b$, $T=0.01 U_0$: (a) domain wall representation; (b) domain representation.}
\label{fig_net_13_neg}
\end{figure}

Here we also extend the results of the main text to the case of the electons' doping which accumulate into interstitials rather than voids (Fig. \ref{fig_net_13_neg}).
For negative doping we still observe the charge fractionalization phenomena and qualitatively similar to the case of positive doping structures. The wall crossings are still favorable (thus no ``stripe phase''), but connection of 4 walls is also favorable as of 3 walls, which is presumably governed by the short-range physics. Because of this, visually the pictures look differently.

For higher $l_s$, the long-range part of the Coulomb interaction will become progressively more important and will favor walls' 3-crossings, with $120^{\mathrm{o}}$ angle between them because of their repulsion, at least for sparse walls' networks (when the walls' structure characteristic period is $\gg l_s$). Thus we expect that the metastable states with the lowest lying energies will represent the ``irregular honeycomb network'' \cite{Villain:1980} as observed for the positive doping.
